\documentclass[gmd, manuscript]{copernicus}

\begin{document}

\nolinenumbers

\title{A new process-based vertical advection/diffusion theoretical model of ocean heat uptake}


\Author[1]{R\'emi}{Tailleux}
\Author[1,2]{Antoine}{Hochet}
\Author[1]{David}{Ferreira}
\Author[1,3]{Till}{Kuhlbrodt}
\Author[1,4]{Jonathan}{Gregory}

\affil[1]{Department of Meteorology, University of Reading, Earley Gate, PO Box 243, Reading, RG6 6BB, United Kingdom}
\affil[2]{Scripps Institution of Oceanography, University of California San Diego}
\affil[3]{NCAS, University of Reading, Earley Gate, PO Box 243, Reading, RG6 6BB, United Kingdom}
\affil[4]{Met Office Hadley Centre, Exeter, United Kingdom}


\runningtitle{New process-based theoretical model of ocean heat uptake}

\runningauthor{Tailleux et al.}

\correspondence{R\'emi Tailleux (R.G.J.Tailleux@reading.ac.uk)}

\received{}
\pubdiscuss{} 
\revised{}
\accepted{}
\published{}


\firstpage{1}

\maketitle

\begin{abstract}
The vertical upwelling/diffusion model (VUDM) has historically played a key role in shaping our ideas about how the heat balance is achieved in the ocean. Its has been and is still widely used in many applications ranging from the estimation of transfer coefficients to the parameterisation of ocean heat uptake in Simple Climate Models (SCMs). Its conceptual value as a realistic theoretical model of the ocean heat balance has become increasingly unclear over the years however, because: 1) the different ways in which upwelling has been linked to high-latitude deep water formation and downgradient diffusion linked to vertical/diapycnal mixing have remained imprecise and somewhat ad-hoc so far; 2) other effects such as isopycnal mixing, density-compensated temperature anomalies, meso-scale eddy-induced advection and the depth-varying ocean area have all be demonstrated to affect actual ocean heat uptake as well, but their incorporation into existing VUDM frameworks has been problematic. In this paper, a new process-based vertical advection/diffusion theoretical model of ocean heat uptake is constructed that resolve all above difficulties. This new model is obtained by coarse-graining the full three-dimensional advection/diffusion for potential temperature carried by ocean climate models, by using the same isopycnal analysis as in the theory of ocean water masses. The resulting model describes the temporal evolution of the isopycnally-averaged thickness-weighted potential temperature in terms of an effective velocity that depends uniquely on the surface heating conditionally integrated in density classes, an effective diapycnal diffusivity controlled by isoneutral and dianeutral mixing, and an additional term linked to the meridional transport of density-compensated temperature anomalies by the diabatic residual overturning circulation. 
\end{abstract}

\introduction  

Ocean heat uptake limits the overall rate of global warming caused by increasing greenhouse gas emissions. Understanding the physical processes that control it is therefore a key issue in the study of climate change. An important difficulty impeding progress, however, stems from that most of the key physical processes controlling ocean heat uptake --- such as the turbulent molecular diffusion of heat and salt, deep water formation, meso-scale eddy-induced advection --- occur at scales too small to be resolved directly in numerical ocean models. As a result, the latter need to be parameterised, but how best to do this is still an open question that has occupied fluid dynamicists, oceanographers and ocean modellers for nearly a century. Although some important progress has been achieved over the past decades, such as the introduction of a meso-scale eddy parameterisation and the use of rotated diffusion, important model biases in the simulated temperature and salinity fields remain that suggest that a full understanding of ocean heat uptake remains limited. 

To discuss the relative importance of the various physical issues affecting our understanding ocean heat uptake, conceptual models are useful. Various types have been developed over the years, such as the two-layer model (Gregory, Held) or that based on linear response theory  (Good et al.). One of the most widely discussed, however, and the one that is the focus of the present paper is probably the vertical advection/diffusion model (VADM) considered by \cite{hoffert1980role,munk1966abyssal}, which can be written under the form:
\begin{equation}
    \frac{\partial \overline{\theta}}{\partial t} + w_{\rm eff} \frac{\partial \overline{\theta}}{\partial z} = \frac{\partial}{\partial z} \left ( K_{\rm eff} \frac{\partial \overline{\theta}}{\partial z} \right ) ,
    \label{VADM}
\end{equation}
where $K_{\rm eff}$ and $w_{\rm eff}$ represent an effective vertical advection and diffusivity respectively. In Munk's paper, $w_{\rm eff}$ was envisioned as being associated with the broad upwelling thought to occur in response to high-latitude deep water formation, as in Stommel and Aarons conceptual model of the abyssal circulation. In such a view, a relevant scaling for the effective vertical velocity is $w_{\rm eff} \propto S/A$, where $S$ measures the rate of deep water being formed, and $A$ the area of the ocean across which the upwelling takes place.  
In Munk (1966), this model was considered in a particular region of the Pacific Ocean and regarded as an approximation of the full advection/diffusion equation in a region where the isothermal surfaces could be regarded as approximately flat, and of negligible lateral advection. From a global inversion consideration, Munk concluded that a diffusivity $K_{\rm eff} = O(10^{-4}m^2/s)$ was required in order to balance the rate of cooling of the deep ocean due to high-latitude water masses formation. Pelagic measurements reveal that actual diffusivities in the thermocline were about one order of magnitude less. 

In simple climate models such as MAGICC, e.g.,\cite{meinshausen2011emulating} it has been common to constrain the effective advection and diffusion entering (\ref{VADM}) by means of a {\em behavioural} calibration, that is, one that constrain the parameter so that the model emulate a given target behaviour. SCMs represent a key tool for policy advice on mitigation strategies, and there is considerable interest in understanding the limits of such models, or what determine the range of validity of (\ref{VADM}) or models similar to that. 

So far, the mean temperature $\overline{\theta}(z,t)$ appearing in (\ref{VADM}) has been commonly interpreted as the horizontally-averaged (potential) temperature, e.g., \cite{wolfe2008vertical,Huber:2015fu}, as assumed for instance in simple climate models such as MAGICC, e.g., \cite{meinshausen2008emulating}. In this interpretation, the effective vertical advection is related to the vertical heat transport associated with the horizontal correlation between vertical velocity and heat, while the effective diffusivity is related to the diapycnal and isopycnal diffusive heat fluxes, with the latter exhibiting a tendency to occasionally make $K_{\rm eff}$ small if not slightly negative. At the surface, only the net surface-integrated heat flux plays a role, all information relative to geographical distribution of the surface heat fluxes is lost. 

* Discussion of what ocean heat uptake in terms of physical processes (Kuhlbrodet and Gregory)

* Importance of the Southern Ocean, residual circulation, Zanna and Marshall idealised study.

The main aim of this paper is to propose a rigorous construction of (\ref{VADM}) based on an isopycnal average of the full three-dimensional advection/diffusion for potential temperature used in OGCMs.

\section{Isopycnal averaging of the three-dimensional temperature and salinity equations}

\subsection{Model assumptions and definitions}

Because the main aim of SCMs is to emulate the behaviour of comprehensive coupled climate models, we take as our starting point the full three-dimensional evolution equations for potential temperature and salinity carried out in such models, which we take to be of the form: 
\begin{equation}
    \frac{\partial \theta}{\partial t} = - {\bf v}_{res}\cdot \nabla \theta + \nabla \left ( {\bf K} \nabla \theta \right ) , 
    \label{3d_advection_diffusion}
\end{equation}
\begin{equation}
     \frac{\partial S}{\partial t} = -{\bf v}_{res} \cdot \nabla S + \nabla \left ( {\bf K} \nabla S \right ) ,
     \label{3d_advection_diffusion_salt}
\end{equation}
where ${\bf K} = K_i ({\bf I} - {\bf d}{\bf d}^T ) + K_d {\bf d} {\bf d}^T$ denotes the (neutral) rotated diffusion tensor, ${\bf d} = {\bf N}/|{\bf N}|$ the normalised neutral vector, while $K_i$ and $K_d$ denote the isoneutral and dianeutral diffusivities. The residual velocity ${\bf v}_{res}$ is defined as the sum of the Eulerian mean velocity ${\bf v}$, solution of the prognostic equations for momentum, and of the parameterised meso-scale eddy-induced velocity ${\bf v}_{gm}$. The neutral vector is defined by ${\bf N} = g (\alpha \nabla \nabla \theta - \beta \nabla S)$, where $g$ is the acceleration of gravity, $\alpha$ is the thermal expansion coefficient, and $\beta$ is the haline contraction coefficient. Note that rotated diffusion only pertains to mixing in the ocean interior, as mixing in the mixed layer is normally achieved by horizontal mixing. 

\subsection{Construction of the reference density and temperature profiles}

In order to coarse-grain (\ref{3d_advection_diffusion}-\ref{3d_advection_diffusion_salt}) using an isopycnal average, we need to identify a suitable purely material density variable $\gamma(S,\theta)$ (where $S$ and $\theta$ can be replaced by absolute salinity and Conservative Temperature for ocean models using such variables). Ideally, $\gamma$ should be as free as feasible of vertical inversions in order to facilitate the passage from vertical to density coordinates, although this is not essential. For a stably stratified ocean, let us note that if it were possible to construct an exactly neutral density variable, the latter would  be naturally free from vertical inversions since its vertical gradient would then be proportional to $N^2>0$. As is well known, however, this is forbidden by the coupling between thermobaricity and density-compensated anomalies, e.g. \cite{Tailleux:2015tn,tailleuxdir}, which means that the best that can be achieved is a variable maximising neutrality somehow \citep{Jackett:1997ff,tailleuxdir}. In this regards, the best material density variable appears to be a function of Lorenz reference density $\rho_{LZ}(S,\theta) = \rho(S,\theta,p_r^{LZ}(S,\theta))$, that is the potential density referenced to Lorenz reference pressure $p_r^{LZ}(S,\theta)$, that is the pressure would have in a notional state of rest obtained by means of an adiabatic and isohaline re-arrangement of the actual state \citep{tailleuxdir}, but has yet to be fully explored.  
\par
Regardless of the particular material density variable $\gamma(S,\theta)$ chosen to perform the sought-for isopycnal averaging, an essential first step is the construction of a depth-dependent reference state $\gamma_r(z,t)$ and its inverse function $z_r(\gamma,t)$, which is required to switch from depth-coordinates to density coordinates and conversely. Once both functions have been constructed, it is possible to write the isopycnal surface $\gamma={\rm constant}$ equivalently in the form: $z=\zeta(x,y,z_r,t) = \zeta(x,y,z_r(\gamma,t),t) = \hat{\zeta}(x,y,\gamma,t)$. In other words, when working in isopycnal coordinates, we can regard any function of space and time as either a function of $(x,y,z_r,t)$ or $(x,y,\gamma,t)$, with a hat being used in the latter case. In the following, we will tend to use the first option, which tends to make derivations simpler. Note here that this is in contrast with the literature, which tends to prefer the use of $\gamma$ rather than $z_r$. 
\par
In order to construct our reference state, we map the volume of water $\hat{V}(\gamma,t) = V(z_r)$ of water masses with density lower than $\gamma$ with its p.d.f. distribution mapped into physical space, as done for Lorenz reference state by \cite{Saenz:2015vo}. Mathematically, this is equivalent to defining the reference depth $z_r(\gamma,t)$ trough the following equation:
\begin{equation}
    V_r = V(z_r) 
    = \int_{z_r}^0 A(z) \,{\rm d}z 
    = \int_x \int_y \int_{\zeta(x,y,z_r,t)}^0 {\rm d}z \,{\rm d}y \,{\rm d}x = - \int_x \int_y \zeta(x,y,z_r,t) \,{\rm d}x{\rm d}y ,
    \label{reference_volume}
\end{equation}
where $A(z)$ is the area of the ocean at depth $z$.
Note that $\zeta(x,y,z_r,t)=\hat{\zeta}(x,y,\gamma,t)$ vanishes at the outcropping locations. Likewise, we can define the isopycnally-averaged potential temperature through the following relationship
\begin{equation}
     \int_{V_r} \theta({\bf x},t)\,{\rm d}V 
     = \int_S \int_{\zeta(x,y,z_r,t)}^0 \theta(x,y,z,t) 
     {\rm d}z {\rm d}x {\rm d}y = \int_{z_r}^0 
     A(z) \theta_r(z,t)\,{\rm d}z .
     \label{theta_definition}
\end{equation}
In order to obtain an explicit expression for $\theta_r(z,t)$, it is useful to differentiate Eqs. (\ref{theta_definition}) and (\ref{reference_volume}) with respect to $z_r$, respectively:
\begin{equation}
    \int_S \theta(x,y,\zeta(x,y,z_r,t),t) \frac{\partial \zeta}{\partial z_r}\,{\rm d}x{\rm d}y = A(z_r) \theta_r(z_r,t) 
\end{equation}
\begin{equation}
    A(z_r) = \int_S \frac{\partial \zeta}{\partial z_r}\,{\rm d}x {\rm d}y 
\end{equation}
Note that although the outcropping latitudes depend on $\gamma$, their derivatives does not enter the final expressions because $\zeta=0$ at the outcropping latitudes by definition. By combining these two expressions, the following operational definition of $\theta_r$ is obtained:
\begin{equation}
  \theta_r(z_r,t) = \left ( \int_S \frac{\partial \zeta}{\partial z_r}\,{\rm d}x {\rm d}y \right )^{-1} \int_S \frac{\partial \zeta}{\partial z_r} \theta(x,y,\zeta(x,y,z_r,t),t) \,{\rm d}x {\rm d}y . 
\end{equation}
This expression shows that $\theta_r$ should be defined as a form of thickness weighted averaged potential temperature, similarly as considered by Young (2012),  where the thickness is defined here by $h = (\partial \zeta/\partial z_r){\rm d}z_r = (\partial \hat{\zeta}/\partial \gamma) \,{\rm d}\gamma$. Note that inverting $\theta_r(z_r,t)$ yields the reference density profile for density in the form $\gamma = \gamma_r(z_r,t)$.
\par 
Also useful in the following are the following results:
\begin{equation}
    \frac{\partial V_r}{\partial t} = - \int_{S} \frac{\partial \zeta}{\partial t}\,{\rm d}x{\rm d}y = - A(z_r) \frac{\partial z_r}{\partial t} = 0 ,
    \label{dvrdt_zero}
\end{equation}
\begin{equation}
    \frac{\partial V_r}{\partial z_r} = - A(z_r) = - \int_{S} \frac{\partial \zeta}{\partial z_r}\,{\rm d}x{\rm d}y .
    \label{dvrdzr}
\end{equation}

\subsection{Evolution equation for the reference temperature profile}

In order to obtain an evolution equation for $\theta_r(z_r,t)$, we proceed in two steps. The first step consists in integrating Eq. (\ref{3d_advection_diffusion}) over the volume of water $V(z_r)$ as follows:
\begin{equation}
     \underbrace{\int_{V_r} \frac{\partial \theta}{\partial t} \,{\rm d}V}_{T_1(z_r,t)} =  \underbrace{-\int_{V_r} \nabla \cdot ( \theta {\bf v}_{res} ) \,{\rm d}V}_{T_2(z_r,t)} + \underbrace{\int_{V_r} \nabla \cdot ({\bf K} \nabla \theta) \,{\rm d}V}_{T_3(z_r,t)} . 
     \label{integrated_advection_equation}
\end{equation}
This integration provides a number of terms that each depends only on $z_r$ and time. As is made clear further on, an evolution equation for $\theta_r(z_r,t)$ can then be obtained by differentiating each term with respect to $z_r$. A theoretical analysis of each of the term is described in Appendix A. For example, we show in the Appendix that 
\begin{equation}
    \frac{\partial T_1}{\partial z_r} = -A(z_r) \frac{\partial \theta_r}{\partial z_r} + \frac{\partial}{\partial z_r} \int_S \left ( \theta - \theta_r \right ) \frac{\partial \zeta}{\partial z_r}\,{\rm d}x {\rm d}y .
\end{equation}
After some manipulation, based on the results of the Appendix, we can write the following evolution equation:
\begin{equation}
     \frac{\partial \theta_r}{\partial t} + w_{\rm eff}^{\theta} \frac{\partial \theta_r}{\partial z} = \frac{1}{A(z_r)} \frac{\partial}{\partial z_r} \left ( A(z_r) K_{\rm eff}^{\rm mixing} \frac{\partial \theta_r}{\partial z_r} \right ) + C_{\theta}(z_r,t)
\end{equation}
where the term $C_{\theta}(z_r,t)$ is given by
\begin{equation}
     C_{\theta}(z_r,t) = -\frac{1}{A(z_r)} \frac{\partial}{\partial z_r} \left [ 
     \int_S (\theta-\theta_r) \frac{D_{\rm res}z_r}{Dt} \frac{\partial \zeta}{\partial z_r} {\rm d}x {\rm d}y + \int_S [{\bf K}\nabla (\theta-\theta_r)] \cdot {\bf n}{\rm d}S \right ]
\end{equation} 
while the effective vertical velocity is given by:
\begin{equation}
    w_{\rm eff} = \frac{1}{A(z_r)} \left ( \frac{\partial \theta_r}{\partial z_r} \right )^{-1} \left [ \frac{\partial {\cal H}}{\partial z_r} + \frac{\partial T_4}{\partial z_r} \right ] .  
\end{equation}
We expect that ${\cal H}(\gamma,t)$ should be initially zero for $\gamma_{min}$ and then increases with $\gamma$ for low values of $\gamma$, then reach a maximum, before eventually decreasing, so that for values typical of the deep ocean, one should have $\partial {\cal H}/\partial \gamma < 0$. I believe that we will need to include some hypsometric effect in the definition of the problem.

\section{Isopycnal-averaging of the equation for salinity and density}

\subsection{Salinity}
As described above, an important feature of the isopycnally-averaged equation for potential temperature is the existence of a previously overlooked contribution from density-compensated temperature anomalies. The existence of such a term suggests some form of coupling between temperature and salinity. By developing the same approach as for potential temperature, we can similarly derive an evolution equation for the isopycnally-averaged thickness-weighted salinity profile $S_r(z,t)$ a similar equation as that for $\theta_r(z,t)$, which takes the form:

\begin{equation}
    \frac{\partial S_r}{\partial t} + w_{\rm eff}^S \frac{\partial S_r}{\partial z_r}
    = \frac{\partial}{\partial z_r} \left ( K_{\rm eff}^{\rm mixing} \frac{\partial S_r}{\partial z_r} \right ) + C_S(z_r,t)
\end{equation}
The effective diapycnal mixing diffusivity $K_{\rm eff}^{\rm mixing}$ is the same as for the isopycnal mean potential temperature, as well as for the material density variable $\gamma$ defined below. The effective velocity differs from that for potential temperature and is given by:
\begin{equation}
    w_{\rm eff}^S \frac{\partial S_r}{\partial z_r} = \frac{1}{A(z_r)} \frac{\partial {\cal H}_S}{\partial z_r} 
\end{equation}
This expression is more questionable than for temperature, because $\partial S_r/\partial z_r$ is not necessarily a monotonic function of $z_r$. As a result, it might be more appropriate to avoid using effective advection altogether. 
\par
The last term is given by
\begin{equation}
   C_S(z_r,t) = \frac{1}{A(z_r)} \left [ \frac{\partial}{\partial z_r} \int_S 
   (S-S_r) \frac{D_{\rm res}z_r}{Dt} \frac{\partial \zeta}{\partial z_r} {\rm d}x {\rm d}y
   + \int_S [{\bf K}\nabla (S-S_r)] \cdot {\bf n}\,{\rm d}S \right ] 
\end{equation}
and this time involves density-compensated temperature anomalies.

\subsection{Isopycnal averaging of the equation for $\gamma$}

Note that the evolution equation for the reference temperature is also related to that of the reference density profile. Indeed, we have by construction $\theta_r(z_r,t) = \theta_r(z_r(\gamma,t),t) = \theta_r(\gamma,t)$, so we can in principle deduce the evolution equation of $\theta_r$ from that of the reference density profile. The density, unlike temperature, is directly affected by the nonlinearities of the equation of state. We can hope, therefore, to link the term related to the overturning circulation directly to the effects of cabelling hopefully. 
\begin{equation}
   \frac{D_{res}\gamma}{Dt} = \nabla \cdot ( {\bf K} \nabla \gamma) - NL + \frac{\partial F_{\gamma}}{\partial z}
\end{equation}
where the nonlinear term is given by
$$
    NL = ({\bf K}\nabla \theta )\cdot \nabla \gamma_{\theta} 
    + ( {\bf K} \nabla S ) \cdot \nabla \gamma_S 
$$
\begin{equation}
   = \gamma_{\theta \theta} \nabla \theta \cdot ( {\bf K} \nabla \theta ) 
   + 2 \gamma_{S\theta} \nabla S \cdot ( {\bf K} \nabla \theta ) 
   + \gamma_{SS} \nabla S \cdot ( {\bf K} \nabla S ) . 
\end{equation}
The evolution equation now reads
\begin{equation}
   \frac{\partial \gamma_r}{\partial t} + w_{\rm eff}^{\gamma} \frac{\partial \gamma_r}{\partial z_r} = \frac{1}{A(z_r)}\frac{\partial}{\partial z_r} \left ( A(z_r) K_{\rm eff}^{\rm mixing} \frac{\partial \gamma_r}{\partial z_r} \right ) + C_{\gamma}(z_r,t)
\end{equation}
\begin{equation}
   \frac{\partial \gamma_r}{\partial t} + w_{\gamma} \frac{\partial \gamma_r}{\partial z_r} = \frac{1}{A(z_r)}\frac{\partial}{\partial z_r} \left ( A(z_r) K_{\rm eff}^{mixing} \frac{\partial \gamma_r}{\partial z_r} \right )  - NL 
\end{equation}
Now, it is important to remark that $w_{\gamma}$ is in general different from the effective diffusivity defined for potential temperature. Now, if we regard $\gamma$ as a function of $\theta$, using $\gamma = \gamma_r(z_r,t) = \gamma_r(z_r(\theta_r,t),t) = \tilde{\gamma}_r(\theta,t)$, we can differentiate
$$
   \frac{\partial \tilde{\gamma}}{\partial \theta_r} \frac{\partial \theta_r}{\partial t} 
  + w_{\gamma}\frac{\partial \gamma}{\partial \theta} \frac{\partial \theta_r}{\partial z_r} = \frac{1}{A(z_r)} \frac{\partial}{\partial z_r} \left ( A(z_r) K_{\rm eff}^{mixing} \frac{\partial \gamma}{\partial \theta} \frac{\partial \theta_r}{\partial z_r} \right ) - NL 
$$

\section{Estimates of the different terms}

\subsection{Effective velocity}

\subsection{On the inversion of effective mixing}

In a steady-state, we can identify the value of $K_{\rm eff}$ assuming that $w_{\rm eff}$ is known. We have the equation
$$
   K_{\rm eff} A(z_r) \frac{\partial \theta_r}{\partial z_r} 
   = {\cal H}(z_r,t) 
$$
or
$$
    \overline{K}_{\rm loc}(z_r) = \frac{1}{A_{sk}(z_r)}
    {\cal H} (z_r,t) 
$$

\conclusions  

In this paper, we revisited the construction of the classical vertical/advection model for the ocean heat balance whose use in oceanography can be traced at least as far back to Wyrtki (1961). To that end, we proposed a new process-based rigorous construction of the vertical advection/diffusion model that resolves a number of difficulties associated with the traditional model. The new model is formulated for the isopycnally-averaged thickness weighted potential temperature, and is rooted in the classical theory of water masses. It describes the evolution of the mean isopycnal temperature as a competition between three different terms: 1) surface modification whenever isopycnal surfaces outcrop, which is expressed in terms of an effective velocity that usually corresponds to upwelling in the deep ocean, and downwelling in the upper ocean; 2) effective downgradient diapycnal mixing causing downward diffusion of heat, which is achieved through both dianeutral and isoneutral mixing, and which depends explicitly on the depth-varying ocean area; 3) a new term related to the meridional transport of density-compensated temperature anomalies by the diabatic residual meridional overturning circulation. Physically, this term is expected to act in an anti-diffusive manner for the present day ocean, in the sense that it would contribute to reduce the overall effective diffusivity if it were combined with term 2). The anti-diffusive behaviour is due to the North-South contrast in density-compensated temperature anomalies being positive in the present day ocean.

Because our new approach provides analytical expressions linking all the terms to quantities that can in principle be diagnosed from numerical model outputs, it will be of interest to use the newly developed model to interpret ocean heat uptake in a wide range of climate change experiments, in order to understand how it could be adapted for use in simple climate models such as MAGICC. This will be discussed in a forthcoming publication.

\appendix
\section{Theoretical determination of the terms $T_1$ to $T_4$}   
\subsection{The temporal derivative term $T_1(z_r,t)$}

The term $T_1(z_r,t)$ is related to the temporal variations of the volume integral of the potential temperature. The main aim is to transform it into a volume interal for $\theta_r(z_r,t)$. To that end, straightforward algebra yields:
$$
   T_1(z_r,t) = \int_{V(z_r)} \frac{\partial \theta}{\partial t}{\rm d}V = \frac{\partial}{\partial t} \int_{V(z_r)}\theta\,{\rm d}V + \int_S \theta(x,y,\zeta,t) \frac{\partial \zeta}{\partial t} {\rm d}x {\rm d}y 
$$
$$
    = \frac{\partial}{\partial t} \int_{z_r}^0 A(z) \theta_r(z,t){\rm d}z 
    + \int_S \theta(x,y,\zeta,t) \frac{\partial \zeta}{\partial t} {\rm d}x {\rm d}y 
$$
$$
   = \int_{z_r}^0 A(z) \frac{\partial \theta_r}{\partial t}(z,t){\rm d}z 
 + \int_S \theta(x,y,\zeta,t) \frac{\partial \zeta}{\partial t} {\rm d}x {\rm d}y 
$$
\begin{equation}
   = \int_{z_r}^0 A(z) \frac{\partial \theta_r}{\partial t}(z,t){\rm d}z + \int_{S} \left [ \theta(x,y,\zeta,t) -\theta_r(z_r,t) \right ]\frac{\partial \zeta}{\partial t} {\rm d}x {\rm d}y .
   \label{T1_term}
\end{equation}
The first line uses integration by parts; second line was obtained from the first one by replacing the volume integral of $\theta$ by a volume integral for $\theta_r(z_r,t)$; the third line was obtained by interverting the integration and time-derivative; the fourth line incorporated the term $\theta_r(z_r,t)$ by making use of the result (\ref{dvrdt_zero}). 
\par
In order to clarify the physics of the last term in (\ref{T1_term}), it is useful to link $\theta(x,y,\zeta,t)$ to the reference temperature profile as follows:
$$
   \theta(x,y,\zeta(x,y,z_r,t)) = \theta_r(z_r + \delta \zeta,t) , 
$$
which provides an implicit definition for the quantity $\delta \zeta$. This makes it possible to write:
$$
   \theta(x,y,\zeta,t) - \theta_r(z_r,t) = \theta_r(z_r+\delta \zeta,t) - \theta_r(z_r,t) = \int_{z_r}^{z_r + \delta \zeta} \frac{\partial \theta_r}{\partial z}(z',t)\,{\rm d}z'
   = \frac{\partial \theta_r}{\partial z_r}(z_r,t) \delta \zeta^{\ast}
$$
where the latter expression can be regarded as the definition of the equivalent displacement $\delta \zeta^{\ast}$. This makes it possible, therefore, to rewrite 
$$
   \int_{S} \left [ \theta(x,y,\zeta,t) -\theta_r(z_r,t) \right ]\frac{\partial \zeta}{\partial t} {\rm d}x {\rm d}y = \frac{\partial \theta_r}{\partial z_r}(z_r,t)\int_{S} \delta \zeta^{\ast} \frac{\partial \zeta}{\partial t} \,{\rm d}x {\rm d}y 
   = K_t(z_r,t) A(z_r) \frac{\partial \theta_r}{\partial z_r}(z_r,t) 
$$
where the latter expression can be regarded as the definition of the pseudo diffusivity $K_t$. 
Now, differentiating (\ref{T1_term}) with respect to $z_r$ yields:
\begin{equation}
   \frac{\partial T_1}{\partial z_r} = -A(z_r) \frac{\partial \theta_r}{\partial t} 
  + \frac{\partial}{\partial z_r} \left ( K_t(z_r,t) A(z_r) \frac{\partial \theta_r}{\partial z_r} \right ) .
\end{equation}
We expect the last term to have a negligible contribution to the problem, and it will not be considered further. 

\subsection{The residual advection term $T_2(\gamma,t)$}

The term $T_2$ is related to the cross-isopycnal advection of $\gamma$-compensated temperature anomalies by the residual velocity ${\bf v}_{res}$. Mathematically, it can be written as the following surface integral along $\gamma={\rm constant}$, viz.,  
\begin{equation}
  T_2(z_r,t) = -\int_{V(z_r)} {\bf v}_{res}\cdot \nabla \theta \,{\rm d}V = -\int_{\partial V_r} \theta {\bf v}_{res} \cdot {\bf n} {\rm d}S = -\int_{\partial V_r} [\theta - \theta_r(z_r,t)]{\bf v}_{res}\cdot {\bf n}{\rm d}S , 
\end{equation}
where ${\bf n} = \nabla \gamma/|\nabla \gamma|= -\nabla z_r/|\nabla z_r|$ is the unit outward normal vector to the $\gamma={\rm constant}$ surface (or equivalently, the constant $z_r$ surface). Physically, this term is related to the spiciness transport by the diabatic component of the meridional overturning circulation. Indeed, note that we have ${\bf v}_{res}\cdot {\bf n} = {\bf v}_{res} \cdot \nabla \gamma/|\nabla \gamma| = |\nabla \gamma|^{-1} \left ( D_{res}\gamma/Dt - \partial \gamma/\partial t \right )$ by definition, which is therefore entirely controlled by diabatic effects in a steady-state $\partial \gamma/\partial t \approx 0$.
\par
In order to estimate its contribution to the overall heat balance of the bowl of water with density less than $\gamma$, let us consider the particular case where the diabatic residual circulation 
corresponds to northward flow both in the Northern and Southern Hemispheres, as envisioned in the adibatic theories of the overturning circulation, e.g., \cite{Wolfe:2009rp}. Accordingly, let us assume that the light-to-dense water mass conversion at temperature $\theta=\theta^N$ in the North is compensated by a light-to-dense water mass conversion in the southern surface region of the ACC at temperature $\theta^S$, thus allowing us to write
$$
    T_2 \approx - (\theta^N - \theta^S) \times {\rm MOC} < 0 ,
$$
where ${\rm MOC}$ is the positive volume transport through $\gamma$ in the North. Because in the current ocean, density compensated temperature anomalies on a given density surface tends to be larger in the North Atlantic than in the South Atlantic, we expect $\theta^N-\theta^S$ to be positive, and hence $T_2$ to be strictly negative, and hence to represent a cooling tendency for $\theta_r$. It follows that if we were to combine $T_2$ with the effective advection term, it would contribute to increase the effective upwelling in the deep ocean. Conversely, if we were to combine this term with the effective diffusivity, it would act as an antidiffusive contribution that would reduce $K_{\rm eff}^{mixing}$.
\par
One way to formalise the latter idea is by introducing an equivalent diffusivity $K_{res}$ defined by:
$$
    - \int_{\partial V_r} [\theta - \theta_r(z_r,t)]\,{\bf v}_{res}\cdot {\bf n}\,{\rm d}S = - \frac{\partial \theta_r}{\partial z_r}(z_r,t) \int_{\partial V_r} \delta \zeta^{\ast} {\bf v}_{res}\cdot {\bf n}\,{\rm d}S = - A(z_r) K_{res}(z_r,t) \frac{\partial \theta_r}{\partial z_r}(z_r,t) 
$$
where the latter expression is the definition of the pseudo diffusivity $K_{res}$, viz., \begin{equation} 
    K_{res} = \frac{1}{A(z_r)} \int_{\partial V_r} \delta \zeta^{\ast} {\bf v}_{res}\cdot {\bf n}\,{\rm d}S 
\end{equation}
As a result, we may write
\begin{equation}
    \frac{\partial T_2}{\partial z_r} = - \frac{\partial}{\partial z_r} \left ( A(z_r) K_{res}(z_r,t) \frac{\partial \theta_r}{\partial z_r} \right ) 
\end{equation}

\subsection{The diffusive term $T_3(z_r,t)$}

Integration of the diffusive term yields:
\begin{equation}
  T_3(z_r,t) = 
  \int_{V_r} \nabla \cdot ({\bf K} \nabla \theta ) \,{\rm d}V = 
  \int_{\partial V_r} ({\bf K} \nabla\theta_r)\cdot {\bf n} {\rm d}S + \underbrace{\int_{\partial V_r} ({\bf K} \nabla \theta' ) \cdot {\bf n} {\rm d}S}_{T_{3,spice}} + \underbrace{\int_{A_r} \frac{Q}{\rho_0 c_p}\,{\rm d}x {\rm d}y}_{{\cal H}(z_r,t)} ,
  \label{t3_term}
\end{equation}
and can be expressed as the sum of three terms, associated respectively with the diffusion of the isopycnal mean temperature $\theta_r$, the diffusion of the $\gamma$-compensated temperature anomaly $\theta'$, and the contribution to the surface flux, where $A_r$ in the last term in Eq. (\ref{t3_term}) is the fraction of the total surface area associated with waters of density less than $\gamma$, and $Q$ is the net surface heat flux, i.e., the sum of the sensible, latent, incoming shortwave, and outgoing longwave. 

Regarding the first term, it is the one that can be expressed in terms of an effective diapycnal diffusivity. To that end, let us first note that the outward unit normal vector can be written ${\bf n} = \nabla \gamma/|\nabla \gamma| = - \nabla z_r/|\nabla z_r|$. As a result, it is possible to write the local diffusive flux of $\theta_r$ by neutral rotated diffusion as follows:
$$
   ( {\bf K} \nabla \theta_r ) \cdot {\bf n} = 
   \frac{\nabla \gamma^T}{|\nabla \gamma|} \left [ K_i ({\bf I} - {\bf d} {\bf d}^T ) + K_d {\bf d} {\bf d}^T \right ] \frac{\partial \theta_r}{\partial z_r} \frac{\partial z_r}{\partial \gamma} \nabla \gamma
$$
$$
   = \underbrace{\left [ K_i \sin^2{({\bf d},\nabla \gamma)} + K_d \cos^2{({\bf d},\nabla \gamma)} \right ]}_{K_{\rm loc}} \frac{\partial \theta_r}{\partial z_r} \frac{\partial z_r}{\partial \gamma} |\nabla \gamma| ,
$$
which implies for the net diffusive flux:
\begin{equation}
   \int_{\partial V_r} ({\bf K}\nabla \theta_r) \cdot {\bf n}\,{\rm d}S
   =  \frac{\partial \theta_r}{\partial z_r} \int_{\partial V_{\gamma}} K_{\rm loc} |\nabla \gamma| \frac{\partial z_r}{\partial \gamma}{\rm d}S 
= -\frac{\partial \theta_r}{\partial z_r} K_{\rm eff}^{\rm mixing} (z_r,t) A(z_r) .
\end{equation}
The definition of effective diffusivity used in that paper is therefore given by 
\begin{equation}
    K_{\rm eff}^{\rm mixing} = -\frac{1}{A(z_r)}\int_{\partial V_{\gamma}} K_{\rm loc} |\nabla \gamma | \frac{\partial z_r}{\partial \gamma} {\rm d}S = -\frac{1}{A(z_r)} \int_{S} K_{\rm loc} |\nabla \gamma|^2 \frac{\partial z_r}{\partial \gamma}\left | \frac{\partial \gamma}{\partial z} \right |^{-1}\, {\rm d}x {\rm d}y  
\end{equation}
by using the fact ${\rm d}S = |\nabla \gamma| |\partial \gamma/\partial z|^{-1} {\rm d}x {\rm d}y = |\nabla z_r||\partial z_r/\partial z|^{-1}$. Equivalently, we can also use the fact that $\gamma = \gamma_r(z_r,t)$ to rewrite $\nabla \gamma = \partial \gamma_r/\partial z_r \nabla z_r$, leading to the following mathematically equivalent expression:
\begin{equation}
      K_{\rm eff}^{\rm mixing} = \frac{1}{A(z_r)}\int_S K_{\rm loc} |\nabla z_r|^2 \frac{\partial \zeta}{\partial z_r}\,{\rm d}x {\rm d}y .
\end{equation}
In order to develop a further physical understanding of $K_{\rm eff}^{\rm mixing}$, it is useful to introduce two kind of areas. The first is the area of the isopycnal surface $\gamma = {\rm constant}$, which by definition is given by:
\begin{equation}
   A_S = \int_{S} {\rm d}S =
   \int_S |\nabla z_r| \frac{\partial \zeta}{\partial z_r}\,{\rm d}x {\rm d}y ,
\end{equation}
by using again the expression linking the local area element ${\rm d}S$ to the projected flat area element ${\rm d}x {\rm d}y$. The second area of interest is the pseudoa area needed to normalise the diffusivity, that is:
\begin{equation}
    A_{SK} = \int_S |\nabla z_r|^2 \frac{\partial \zeta}{\partial z_r}\,{\rm d}x {\rm d}y .
\end{equation}
Now, if we define the area-average diffusivity $\overline{K}_{\rm loc}(z_r,t)$ by the following relation:
\begin{equation}
     \overline{K}_{\rm loc}(z_r,t) = \frac{1}{A_{SK}(z_r,t)} \int_{S} K_{\rm loc} |\nabla z_r|^2| \frac{\partial \zeta}{\partial z_r}\,{\rm d}x {\rm d}y , 
\end{equation}
then it is easily seen that the effective diffusivity $K_{\rm eff}^{\rm mixing}$ is linked to $\overline{K}_{\rm loc}$ by:
\begin{equation}
     K_{\rm eff}^{\rm mixing} = \frac{A_{SK}(z_r,t)}{A(z_r)} \overline{K}_{\rm loc}(z_r,t) .
\end{equation}
The advantage of this approach is that now $\overline{K}_{\rm loc}$ is directly comparable with the diapycnal mixing itself. 



\section{Further treatment of density-compensated temperature anomaly term}

The above derivations include a number of different terms that are all involving the potential temperature anomaly $\theta' = \theta - \theta_r(z_r,t)$ defined along a surface of constant $z_r$, whose equation we chose to define as $z=\zeta(x,y,z_r,t) = \hat{\zeta}(x,y,\gamma,t)$. Now, the different terms involving $\theta'$ are as follows:
$$
    - \int_S [\theta(x,y,\zeta,t)-\theta_r(z_r,t)] \frac{\partial \zeta}{\partial t} {\rm d}x {\rm d}y - \int_{\partial V_r} (\theta-\theta_r) {\bf v}_{res} \cdot {\bf n}{\rm d}S 
$$
Now, we have to remember that ${\bf n} = \nabla \gamma/| \nabla \gamma| = -\nabla z_r/|\nabla z_r|$. It follows that we can write
$$
    {\bf v}_{res} \cdot {\bf n} {\rm d}S = - {\bf v}_{res}\cdot \frac{\nabla z_r}{|\nabla z_r|}
    \frac{|\nabla z_r|}{|\partial z_r/\partial z|} {\rm d}x {\rm d}y 
$$
where the residual advection term $D_{\rm eff}z_r/Dt$ is defined by
\begin{equation}
    \frac{D_{\rm res}z_r}{Dt} = \frac{\partial z_r}{\partial S} \frac{D_{\rm res}S}{Dt} + \frac{\partial z_r}{\partial \theta} \frac{D_{\rm res}}{Dt} 
\end{equation}
and hence depends purely on diabatic effects. 
Now, we will be able to write
$$
  {\bf v}_{res} \cdot \nabla z_r = \frac{\partial z_r}{\partial S} \left [ \dot{S} - \frac{\partial S}{\partial t} \right ] + \frac{\partial z_r}{\partial \theta}\left [ \cdot{\theta} - \frac{\partial \theta}{\partial t} \right ] = \frac{D_{res}z_r}{Dt} - \frac{\partial z_r}{\partial t}  
$$
Combining the two expressions therefore yields
$$
    \int_S (\theta-\theta_r) \frac{D_{res}z_r}{Dt} \frac{\partial \zeta}{\partial z_r} {\rm d}x {\rm d}y
$$

\begin{acknowledgements}
This work was supported by the grant NE/K016083/1 ``Improving simple climate models through a traceable and process-based analysis of ocean heat uptake (INSPECT)" of the UK Natural Environment Research Council (NERC). The data  and software that were used for this study are available upon request to the corresponding author.
\end{acknowledgements}



 \bibliographystyle{copernicus}
 \bibliography{references}

\end{document}